\input amstex
\input psfig.sty
\magnification 1200
\TagsOnRight
\def\qed{\ifhmode\unskip\nobreak\fi\ifmmode\ifinner
\else\hskip5pt\fi\fi\hbox{\hskip5pt\vrule width4pt height6pt
depth1.5pt\hskip1pt}}
\def\stretch {\noalign{\medskip}}
\define \bCp {\bold C^+}
\define \bCpb {\overline{\bold C^+}}
\define \ds {\displaystyle}
\define \bR {\bold R}
\baselineskip 19.5 pt
\parskip 5 pt

\centerline {\bf INVERSE SCATTERING FOR VOWEL ARTICULATION}
\centerline {\bf WITH FREQUENCY-DOMAIN DATA}

\vskip 20 pt \centerline {Tuncay Aktosun}
\vskip -8 pt
\centerline{Department of Mathematics and Statistics}
\vskip -8 pt
\centerline {Mississippi State University}
\vskip -8 pt
\centerline{Mississippi State, MS 39762}

\vskip 15 pt

\noindent {\bf Abstract}:
An inverse scattering problem is analyzed
for vowel articulation in the human vocal tract.
When a unit amplitude, monochromatic, sinusoidal volume velocity
is sent from the glottis towards the lips, various types of
scattering data are used to examine whether the cross sectional
area of the vocal tract can uniquely be determined by each data
set. Among the data sets considered are the absolute value
of the pressure measured at a microphone placed
at some distance from the lips, the pressure at the lips, and the transfer function
 from the glottis to the lips. In case of nonuniqueness, it is indicated
what additional information may be used for the unique determination.

\vskip 15 pt
\noindent {\bf PACS (2003):}  02.30.Zz, 43.72.Ct\hfil
\vskip -8 pt
\par\noindent {\bf Short title:}
Inverse scattering for vowel articulation

\newpage

\noindent {\bf 1. INTRODUCTION}
\vskip 3 pt

The fundamental inverse problem for vowel articulation
is concerned [1-5] with the determination of the geometry of the human
vocal tract from some data. In this paper,
we consider various types of scattering data in the frequency
domain resulting from a unit amplitude, monochromatic, sinusoidal
volume velocity sent from the glottis, and we analyze whether
each data set uniquely determines the shape
of the vocal tract, or else, what additional information may be used for
the unique recovery.

Let us use $x$ to denote the distance from the glottis
and $l$ for the length of the vocal tract.
Hence,
the lips are located at a distance $l$ from the glottis.
Typically, $l$ varies between 14 cm and 20 cm, usually smaller
for children than for adults and smaller
for females than for males [1,2,5].
Even though the vocal tract is not a right cylinder, to a
good approximation it can be treated as one [3,4].

We will let
$A(x)$ denote the cross sectional area as a function of
the distance from the glottis, and we suppose that $A(x)$
is positive on $[0,l].$
Assuming that the propagation is lossless and planar
(these assumptions are known [3,4] to be reasonable), the acoustics
in the vocal tract is governed [1-5] by
the first-order linear system of partial
differential equations
$$\cases A(x)\,p_x(x,t)+\mu\,v_t(x,t)=0,\\
\stretch A(x)\,p_t(x,t)+\mu\, c^2\,v_x(x,t)=0,\endcases\tag 1.1$$
where $t$ is the time variable, the subscripts $x$ and $t$
denote the respective partial derivatives,
$\mu$ is the air density, $c$ is the speed of sound,
$v(x,t)$ is the volume velocity of the air flow, and $p(x,t)$ is the
pressure at location $x$ and at time $t.$

The volume
velocity is equal to the product of the cross sectional area with the
average velocity of the air molecules crossing that area. The
pressure is the force per unit cross sectional area and is exerted by the
moving air molecules. The air density at room temperature is
$\mu=1.2\times 10^{-3}$ gm/cm$^3.$ The speed of sound varies
slightly with temperature, and $c=3.43\times 10^4$ cm/sec in air at
room temperature. In our analysis of the inverse problem,
we assume that the values of $\mu$ and $c$ are already known.
There is no loss of generality to start the time
at $t=0.$

By using $v_{xt}=v_{tx},$ we can eliminate $v$ in (1.1) and obtain
Webster's horn equation
$$\ds\frac{1}{A(x)}[A(x)\,p_{x}(x,t)]_x-\ds\frac{1}{
c^2}\,p_{tt}(x,t)=0, \qquad x\in(0,l),\quad t>0.$$
Letting
$$\Phi(x,t):=\sqrt{A(x)}\,p(x,t),\tag 1.2$$
we find that $\Phi$ satisfies the plasma-wave equation
$$\Phi_{xx}(x,t)-\ds\frac{1}{c^2}\,\Phi_{tt}(x,t)=Q(x)
\,\Phi(x,t),\qquad x\in(0,l),\quad t>0,\tag 1.3$$
where we have defined
$$Q(x):=\ds\frac{[\sqrt{A(x)}]''}{\sqrt{A(x)}},\tag 1.4$$
with the prime denoting
the $x$-derivative.
The quantity $Q$ is called the relative concavity of the vocal
tract or the potential.
Separating the variables as
$$\Phi(x,t):=\psi(k,x)\,e^{ikct},\tag 1.5$$
we find that $\psi(k,x)$ satisfies
the Schr\"odinger equation
$$\psi''(k,x)+k^2\psi(k,x)=
Q(x)\,\psi(k,x),\qquad x\in(0,l).\tag 1.6$$

The frequency $\nu$ is measured in Hertz and related to
$k$ as
$\nu=\ds\frac{kc}{2\pi}.$
Informally, we can
refer to $k$ as the frequency even though the proper
term for $k$ is the angular wavenumber.

In order to recover $A,$ we will consider various types of data for
$k\in\bR^+$ resulting from the glottal
volume velocity $v(0,t)$ given in (4.1).
As our data sets, we consider
the absolute value of the impedance at the lips,
the absolute value of the pressure measured
at a microphone placed at some distance from the lips,
the absolute value of the pressure at the
lips, the absolute value of the transfer function from the glottis to the lips,
the absolute value of the impedance at the glottis,
the absolute value of a Green's function for (1.3) measured at the lips,
and the real or imaginary part of the reflectance at the glottis.

The inverse problem of recovery of $A$ can be analyzed either
as an inverse spectral problem or as an inverse scattering
problem.
In the inverse-spectral formulation,
in addition to a boundary condition at the the glottis such as (2.1),
a boundary condition is also imposed at the lips. The imposition of
the boundary conditions at both ends of the
vocal tract results in standing waves that are related
to an infinite sequence of discrete frequencies.
It was established by Borg [6] that $Q$ can be recovered by using two
such infinite sequences of discrete frequencies corresponding to two sets of boundary
conditions. It then follows [3,4,7-11] that
$A$ can be recovered from two infinite sequences of constants. For example,
such sequences can be chosen as the zeros and poles [7,8] of the input impedance
or the poles and residues [9] of the input impedance.

In the the inverse-scattering formulation,
a boundary condition is imposed at only one end of the
vocal tract---either at the glottis or at the lips.
Then, the measurement of the acoustic data used in
the recovery of $A$ is performed at
the same end or at the opposite end.
If the boundary condition and the measurement occur
at the same end of the tract, the corresponding inverse problem is
usually known as a reflection problem. On the other hand, if the
boundary condition and the measurement occur at different ends, then
we have a transmission problem.
The methods based on the inverse scattering formulation may be applied
either in the time domain or in the frequency domain,
where the data set is a function of $t$ in the former case
and of $k$ in the latter.
We refer the reader to [3,4,12-15]
for some approaches as time-domain reflection problems and to [16]
for an approach as a time-domain transmission problem.
Our approach in this paper is a frequency-domain approach, where the analysis
in Sections~6-10 may be viewed as that for a transmission problem and the analysis
in Sections~11 and 12 may be viewed as that for a reflection
problem.

Our paper is organized as follows. In Section~2
we review some preliminary material
related to the Schr\"odinger equation and introduce
the selfadjoint boundary condition involving
$\cot\alpha$ given in (2.1), the Jost solution
$f,$ the Jost function $F_\alpha,$
and the scattering coefficients $T,$ $L,$ and $R.$
In Section~3 we briefly review the recovery of
$Q,$ $\cot\alpha,$ $F_\alpha,$ $f,$ $T,$ $L,$ and $R$
 from the data $\{|F_\alpha(k)|:\ k\in\bR^+\}.$
In Section~4 we obtain some explicit
expressions for the pressure and the volume velocity
in the vocal tract in terms of $A,$ $f,$ and $F_\alpha,$
and we also show that $\cot\alpha$ appearing in (2.1) is directly related
to the physical parameters $A(0)$ and $A'(0).$
In Section~5 we introduce the relative area $[\eta(x)]^2$ and
express it in terms of the Jost solution,
$\cot\alpha,$ and
the scattering coefficients. In Sections~6-12 we analyze
the recovery of $Q,$ $\eta,$ and $A$ from various data sets. The data set
used in Section~6 is the absolute value of the output impedance at the lips.
In Section~7 it includes
the absolute value of the pressure measured at a microphone placed
at some distance from the lips. In Section~8 it is the absolute value of
the pressure measured at
the lips.
In Section~9 the data set includes the absolute value of the transfer function
 from the glottis to the lips. In Section~10
it is the absolute value of an analog of the Green's function introduced
in [17] for (1.3), in Section~11 the absolute value of the input impedance at the
glottis, and in Section~12 the real or imaginary part of the reflectance.
Finally, in Section~13 we present some examples to illustrate the theoretical
results presented in the earlier sections.

\vskip 10 pt
\noindent {\bf 2. PRELIMINARIES}
\vskip 3 pt

In this section we review the scattering data related to
the potential $Q$ appearing in the Schr\"odinger equation
on the half line $\bR^+$ with the selfadjoint
boundary condition [18-21]
$$\sin\alpha\cdot\varphi'(k,0)+\cos\alpha\cdot\varphi(k,0)=0,\tag 2.1$$
where $\alpha$ is a number in the interval $(0,\pi)$ identifying the
boundary condition at $x=0.$
We can relate the half-line Schr\"odinger equation to
(1.6) by assuming that $Q(x)\equiv 0$
for $x>l.$
Note that the mapping
$\alpha\mapsto\cot\alpha$ is one-to-one and
from $(0,\pi)$ onto $\bR.$

Let $f$ denote the Jost solution [18-23] to the half-line
Schr\"odinger equation. It
is uniquely determined by the asymptotic conditions
$$f(k,x)=e^{ikx}[1+o(1)],\quad f'(k,x)=ik\,
e^{ikx}[1+o(1)],\qquad x\to +\infty.$$
Since $Q$ vanishes when $x>l,$ we have
$$f(k,l)=e^{ikl},\quad f'(k,l)=ik\,e^{ikl}.\tag 2.2$$

The Jost function $F_\alpha$ associated with the half-line
Schr\"odinger equation with the boundary condition (2.1) is defined
as [18-21]
$$F_\alpha(k):=-i[f'(k,0)+\cot\alpha\cdot f(k,0)].\tag 2.3$$
Let us emphasize that the subscript in $F_\alpha$ identifies
the boundary condition at $x=0$ and it does not
indicate any partial derivative.
It is known [18-21] that
$$F_\alpha(-k)=-F_\alpha(k)^\ast,\qquad k\in\bR,\tag 2.4$$
where the asterisk denotes complex conjugation.

We assume that $Q$ is real valued and integrable on
$(0,l)$ and that there are no bound states for the half-line
Schr\"odinger equation with the boundary condition (2.1).
The absence of bound states is equivalent [18-21]
to assuming that $F_\alpha(k)$ has no zeros on $\bold I^+,$
where $\bold
I^+:=i(0,+\infty)$ is the positive imaginary axis in the complex
plane. It is known [19-21] that either $F_\alpha(0)\ne 0$
or $F_\alpha(k)$ has a simple zero at $k=0;$ the former is known
as the generic case and the latter as the
exceptional case. The exceptional case corresponds to the
threshold where the number of bound states can be changed by one
under a small perturbation of the potential.

By using the extension
$Q(x)\equiv 0$ when $x<0,$
we can relate $f(k,0)$ and $f'(k,0)$ to the scattering coefficients
in the full-line Schr\"odinger equation. We have [19,20,22,24]
$$f(k,0)=\displaystyle\frac{1+L(k)}{T(k)},\quad
f'(k,0)=ik\,\displaystyle\frac{1-L(k)}{T(k)},\tag 2.5$$
where $T$ and $L$ denote the transmission
coefficient and the left reflection coefficient, respectively,
associated with $Q.$ The right reflection coefficient $R$ is given by
$$R(k)=-\ds\frac{L(-k)\,T(k)}{T(-k)}.\tag 2.6$$
It is known [19,20,22,24] that
$$T(-k)=T(k)^*, \quad R(-k)=R(k)^*, \quad  L(-k)=L(k)^*, \qquad k \in \bR.
\tag 2.7$$
The absence of bound states for the full-line
Schr\"odinger equation is equivalent [19,20,22,24] for $T(k)$ not to have any
poles on $\bold I^+,$ and this is also equivalent [25] for
$F_{\pi/2}(k)$ not to have any zeros on $\bold I^+.$

\vskip 10 pt
\noindent {\bf 3. RECOVERY OF $Q$ FROM $|F_\alpha|$}
\vskip 3 pt

In the absence of bound states, the fundamental inverse scattering problem
for the half-line Schr\"odinger equation with the selfadjoint boundary condition
(2.1) consists of determining $Q$ and
$\cot\alpha$ from various types of scattering data.
In this section we review the solution to this inverse problem
when the data set is $\{|F_\alpha(k)|:\ k\in\bR^+\}.$

\noindent {\bf Theorem~3.1} {\it Assume that
$Q$ is real valued, measurable, and integrable for $x\in(0,l).$
Then, the data set $\{|F_\alpha(k)|:\ k\in\bR^+\}$
uniquely determines $Q(x)$ for $x\in(0,l)$ and
$\cot\alpha.$ The same data set also uniquely determines
the corresponding Jost solution $f(k,x)$ and the scattering
coefficients $T(k),$ $R(k),$ and $L(k).$}

Below we outline some steps involved in the solution to the inverse problem
stated in Theorem~3.1. As seen from (2.4), $|F_\alpha(k)|$ is an even function of
$k\in\bR,$ and hence the data sets
$\{|F_\alpha(k)|:\ k\in\bR^+\}$
and $\{|F_\alpha(k)|:\ k\in\bR\}$ are equivalent.
By using the data $\{|F_\alpha(k)|:\ k\in\bR\}$
as input in the Gel'fand-Levitan method [18-21], we form the kernel function
$$G_\alpha(x,y):=\ds\frac{1}{\pi}\int_{-\infty}^\infty
dk\,\left[\ds\frac{k^2}{|F_\alpha(k)|^2}-1\right]\left(\cos
kx\right) \left(\cos ky\right),$$
and then solve the Gel'fand-Levitan integral equation
$$h_\alpha(x,y)+G_\alpha(x,y)+\int_0^x dz\,
G_\alpha(y,z)\,h_\alpha(x,z)=0,\qquad 0\le y<x.\tag 3.1$$
The solution to (3.1) is known [19-21] to exist and to be unique.
Once $h_\alpha(x,y)$ is obtained, we recover the potential as
$$Q(x)=2\,\ds\frac{d}{dx}h_\alpha(x,x^-),\qquad x\in(0,l),$$
where $x^-$ indicates that
the limit from the
left must be used in the evaluation.
We also recover the boundary condition as
$$\cot\alpha=-h_\alpha(0,0).$$

Alternatively, we can proceed [21] as follows.
Let
$$\Lambda_\alpha(k):=-1+\ds\frac{k\,f(k,0)}
{F_\alpha(k)},\qquad k\in\bCpb,\tag 3.2$$
where we use $\bCp$ for the upper half complex plane and
$\bCpb$ for $\bCp\cup\bR.$
Then
$$\text{Re}[\Lambda_\alpha(k)]=-1+\ds\frac{k^2}{|F_\alpha(k)|^2},
\qquad k\in\bR.\tag 3.3$$
 From the data
$\{|F_\alpha(k)|:\ k\in\bR\}$
we first  construct the function
$\Lambda_\alpha(k)$ via the Schwarz integral formula as
$$\Lambda_\alpha(k)=\ds\frac{1}{\pi i}\int_{-\infty}^\infty
\ds\frac{dt}{t-k-i0^+}\,
\left[-1+\ds\frac{t^2}{|F_\alpha(t)|^2}\right],\qquad k\in\bCpb,\tag 3.4$$
where the quantity $i0^+$ indicates that the values for
real $k$ should be obtained as limits from $\bCp.$
Next, $F_\alpha(k)$ is obtained from $|F_\alpha(k)|$ by using
$$F_\alpha(k)=k\,
\exp\left(\ds\frac{-1}{\pi i}\int_{-\infty}^\infty dt\,
\ds\frac{\log |t/F_\alpha(t)|}{t-k-i0^+}\right),\qquad k\in\bCpb.\tag 3.5$$
Then, we have
$$f(k,0)=\ds\frac{1}{k}\,F_\alpha(k)\,
[1+\Lambda_\alpha(k)],\qquad k\in\bCpb,\tag 3.6$$
$$f'(k,0)=i\,F_\alpha(k)\left[1+\ds\frac{1+\Lambda_\alpha(k)}{k}\,
\displaystyle\lim_{k\to\infty}\left[k\,
\Lambda_\alpha(k)\right]\right] ,\qquad k\in\bCpb.\tag 3.7$$
$$\cot\alpha=-i\,\displaystyle\lim_{k\to\infty}\left[k\,
\Lambda_\alpha(k)\right],\tag 3.8$$
where the limit in (3.8) can be evaluated in any manner in
$\bCpb.$
Having both $f(k,0)$ and $f'(k,0)$ in hand,
we can construct all the quantities that are relevant
in the scattering theory for the Schr\"odinger equation.
For example,
the scattering coefficients for the full-line
Schr\"odinger equation can be obtained as
$$T(k)=\ds\frac{2ik}{ik\,f(k,0)+f'(k,0)},\quad
L(k)=\ds\frac{ik\,f(k,0)-f'(k,0)}
{ik\,f(k,0)+f'(k,0)},\tag 3.9$$
$$R(k)=\ds\frac{-ik\,f(-k,0)-f'(-k,0)}
{ik\,f(k,0)+f'(k,0)}.\tag 3.10$$
Having obtained such quantities, we can construct
the potential by using any one of the various methods
available [19,20,22,24]. For example, we can use
the Faddeev-Marchenko method [19,20,22,24] and get
$$Q(x)=-2\,\ds\frac{d}{dx}K(x,x^+),\qquad x\in\bR,$$
where
$K(x,y)$ is obtained by solving the (left) Faddeev-Marchenko integral equation
$$K(x,y)+\hat R(x+y)+\int_x^\infty dz\,\hat R(y+z)\,K(x,z)=0,\qquad
-\infty <x<y,\tag 3.11$$
with the kernel
$$\hat R(y):=
\ds\frac{1}{2\pi}
\int_{-\infty}^\infty dk\,R(k)\,e^{iky}.$$
The Jost solution $f(k,x)$ can directly be obtained from
$K(x,y)$ as
$$f(k,x)=e^{ikx}+\ds\int_x^\infty dy\, K(x,y)\,e^{iky}.\tag 3.12$$

Let us remark that that, in order to obtain
$\{\Lambda_\alpha(k):\ k\in\bR\}$ from $\{|F_\alpha(k)|:\ k\in\bR\},$
instead of using (3.4) we can equivalently construct
the real and imaginary parts of $\Lambda_\alpha(k)$ via (3.3) and
$$\text{Im}[\Lambda_\alpha(k)]=\ds\frac{-1}{\pi}\,\text{CPV}\ds\int_{-\infty}^\infty
\ds\frac{dt}{t-k}\left[-1+\ds\frac{t^2}{|F_\alpha(t)|^2}\right],\qquad
k\in\bR,$$
where CPV indicates that the integral must be evaluated as a
Cauchy principal value. Consequently, $\cot\alpha$
can be recovered from the
equivalent form of (3.8) given by
$$\cot\alpha=\displaystyle\lim_{k\to\infty}\left(k\,
\text{Im}[\Lambda_\alpha(k)]\right).$$

\vskip 10 pt
\noindent {\bf 4. PRESSURE AND VOLUME VELOCITY}
\vskip 3 pt

When the vocal tract area function $A$ is known,
via (1.4) we can evaluate the potential
$Q,$ solve the corresponding Schr\"odinger equation, and
obtain the Jost solution $f(k,x).$ In this section, with the
help $f(k,x),$ we
express the pressure and volume velocity corresponding
to the glottal
volume velocity
$$v(0,t)=e^{ikct},\qquad t>0.\tag 4.1$$

It is known [19,20,22] that $f(k,\cdot)$ and $f(-k,\cdot)$
are linearly independent for each $k\in\bCpb\setminus\{0\}.$
Hence, the general solution to (1.6) can be written
as a linear combination of $f(k,\cdot)$ and $f(-k,\cdot).$
 From (1.2), (1.5), and (1.6), we see that
the pressure has the form
$$p(x,t)=P(k,x)
\,e^{ikct},\tag 4.2$$
with
$$P(k,x)=\ds\frac{1}{\sqrt{A(x)}}\,\left[a(k)\,f(-k,x)+b(k)\,f(k,x)\right],\tag 4.3$$
where $a(k)$ and $b(k)$ are coefficients to be determined.
When $x\ge l,$ the pressure $p(x,t)$
should be a wave traveling
outward from the lips and should not
contain the part proportional to $e^{ik(ct+x)}$ traveling
into the mouth.
Thus, with the help of (2.2) we see that
we must have $b(k)\equiv 0$ in (4.3).
Hence, (4.3) is reduced to
$$P(k,x)=\ds\frac{1}{\sqrt{A(x)}}\,a(k)\,f(-k,x)
,\qquad x\in[0,l].\tag 4.4$$
Our aim is to determine the value of
$a(k)$ in terms of the pressure $P(k,l+r)$
measured by a microphone placed at a radial distance
$r$ from the lips.
The relationship between the pressure measured
at the microphone and the volume velocity at the lips is explicitly
known and is given by (cf. (3.1) of [5])
$$p(l+r,t)=\ds\frac{\mu}{4\pi r}\,v_t(l,t-r/c),\tag 4.5$$
where we recall that $\mu$ is the air density and $c$ is the sound speed.

 From (4.1), (4.2), and the first line of (1.1), for
the $x$-derivative of the pressure at the glottis we get
$$P'(k,0)=-\ds\frac{ikc\mu}{A(0)}.\tag 4.6$$
Note that from (4.4) through differentiation we obtain
$$P'(k,0)=\ds\frac{1}{\sqrt{A(0)}}\,a(k)\left[f'(-k,0)-
\ds\frac{A'(0)}{2\,A(0)}\,
f(-k,0)\right],\tag 4.7$$
where we have used
$$\ds\frac{[\sqrt{A(x)}]'}{\sqrt{A(x)}}=\ds\frac{A'(x)}{2\,A(x)}.\tag 4.8$$
A comparison of (4.7) with (2.3) shows that, by choosing
$$\cot\alpha=-
\ds\frac{A'(0)}{2\,A(0)}=-\ds\frac{[\sqrt{A(x)}]'\big|_{x=0}}{\sqrt{A(0)}},\tag 4.9$$
we can write (4.7) as
$$P'(k,0)=\ds\frac{i}{\sqrt{A(0)}}\,a(k)\,
F_{\alpha}(-k).\tag 4.10$$
Comparing (4.6) and (4.10) we get
$$a(k)=-\ds\frac{ck\mu}{\sqrt{A(0)}\,F_\alpha(-k)},$$
and hence we can write (4.4) in the equivalent form
$$P(k,x)=-\ds\frac{c\mu k\,f(-k,x)}
{\sqrt{A(x)}\,\sqrt{A(0)}\,F_\alpha(-k)},\qquad x\in[0,l].\tag 4.11$$
Using (4.2) and (4.11) in the first line of (1.1), we get
$$v_t(x,t)=\ds\frac{ck\,A(x)\,e^{ikct}}{\sqrt{A(0)}\,F_\alpha(-k)}
\,\ds\frac{d}{dx}\left[\ds\frac{f(-k,x)}{\sqrt{A(x)}}\right],\qquad x\in
[0,l],\quad t>0.\tag 4.12$$
In particular, from (4.12) we obtain
$$v_t(l,t-r/c)=\ds\frac{ck\sqrt{A(l)}\,e^{ikc(t-r/c)}}{\sqrt{A(0)}\,F_\alpha(-k)}
\left[f'(-k,l)-\ds\frac{A'(l)}{2\,A(l)}\,f(-k,l)\right],
\qquad t>r/c.\tag 4.13$$
Using (2.2) in (4.13) we have
$$v_t(l,t-r/c)=-\ds\frac{ck\sqrt{A(l)}\,e^{ikc(t-r/c-l/c)}}{\sqrt{A(0)}\,F_\alpha(-k)}
\left[ik+\,\ds\frac{A'(l)}{2\,A(l)}\right],
\qquad t>r/c.\tag 4.14$$
Finally, comparing (4.5) and (4.14) we get
$$p(l+r,t)=-\ds\frac{ck\mu\sqrt{A(l)} \,e^{ikc(t-r/c-l/c)}}{4\pi r\,
\sqrt{A(0)}\,F_\alpha(-k)}
\left[ik+\,\ds\frac{A'(l)}{2\,A(l)}\right],
\qquad t>r/c,\tag 4.15$$
or equivalently, with the help of (4.2), we have
$$F_\alpha(-k)=-\ds\frac{ck\mu\sqrt{A(l)} \,e^{-ik(r+l)}}{4\pi r\,\sqrt{A(0)}\,P(k,l+r)}
\left[ik+\,\ds\frac{A'(l)}{2\,A(l)}\right].\tag 4.16$$
 From (4.16) we can conclude that
$k$ appears as $ik$ in $P(k,l+r)$ and hence
$$P(-k,l+r)=P(k,l+r)^\ast,\qquad k\in\bR.\tag 4.17$$
We emphasize that $P(k,x)$ given in
(4.11) is valid only when $x\in[0,l],$ and hence
$P(k,l+r)$ is not obtained from (4.11) by replacing $x$ by
$l+r$ there.
Finally, we remark that, with the help of
(4.1) and (4.12), we obtain
$$v(x,t)=-\ds\frac{i\,\sqrt{A(x)}\,e^{ikct}}{\sqrt{A(0)}\,F_\alpha(-k)}
\left[f'(-k,x)-\ds\frac{A'(x)}{2\,A(x)}\,f(-k,x)\right],\qquad x\in
[0,l],\quad t>0.\tag 4.18$$

\vskip 10 pt
\noindent {\bf 5. AREA AND RELATIVE AREA}
\vskip 3 pt

Let us view (1.4) as the zero-energy Schr\"odinger equation,
and consider the initial-value problem
$$\cases y''=Q(y)\,y,\qquad x\in(0,l),\\
\stretch
y(0)=\sqrt{A(0)},\quad y'(0)=-\sqrt{A(0)}\,\cot\alpha,\endcases\tag 5.1$$
where $\cot\alpha$ is the quantity in (4.9).
Let $y_1(x)$ and $y_2(x)$ be any two linearly independent
solutions to (5.1) on the interval $[0,l].$ Then, the unique solution to (5.1)
can be written as
$$y(x)=\ds\frac{\sqrt{A(0)}}{[y_1(x);y_2(x)]}\left|
\matrix 0&y_1(x)&y_2(x)\\
\stretch
-1&y_1(0)&y_2(0)\\
\stretch
\cot\alpha&y'_1(0)&y'_2(0)\endmatrix\right|,\tag 5.2$$
where $[F;G]:=FG'-F'G$ denotes the Wronskian. Let us define
$$\eta(x):=\ds\frac{\sqrt{A(x)}}{\sqrt{A(0)}}.\tag 5.3$$
We will refer to
$[\eta(x)]^2$ as the relative area of the vocal tract.
Let us remark that we can write (5.3) in the equivalent form
$$A(x)=A(0)\,[\eta(x)]^2,\qquad x\in[0,l].\tag 5.4$$

Recall that
the Wronskian of any two solutions to the Schr\"odinger
equation is independent of $x,$ and $[y_1(x);y_2(x)]\ne 0$
if and only if $y_1$ and $y_2$ are linearly
independent on $[0,l].$
For example, we can choose
$y_1$ and $y_2$ as the zero-energy Jost solutions
$g_{\text l}(0,x)$ and $g_{\text r}(0,x),$ respectively, for the
full-line Schr\"odinger equation where the potential agrees with $Q(x)$ on
the interval $(0,l),$ is zero when $x<0,$ but is
a real-valued, measurable, integrable function with a finite first moment
when $x>l.$ Let $\tau(k),$ $\ell(k),$ and $\rho(k)$ be
the corresponding transmission coefficient, the left reflection coefficient,
and the right reflection coefficient,
respectively. Generically, we have $\tau(0)=0$ or equivalently
$[g_{\text l}(0,x);g_{\text r}(0,x)]\ne 0.$ In the exceptional case,
we have $\tau(0)\ne 0$ or equivalently
$[g_{\text l}(0,x);g_{\text r}(0,x)]=0.$

In the generic case, using [19,20,22,24]
$$[g_{\text r}(k,x);g_{\text l}(k,x)]=\ds\frac{2ik}{\tau(k)},\tag 5.5$$
$$g_r(k,x)=\ds\frac{g_{\text l}(-k,x)+\rho(k)\,g_{\text l}(k,x)}{\tau(k)},$$
$$g_{\text r}(0,0)=1,\quad g'_{\text r}(0,0)=0,$$
we can write (5.2) as
$$\eta(x)=
\left|
\matrix 0&-\ds\frac{i}{2}\,\dot \tau(0)\,
g_{\text l}(0,x)&i\,\dot g_{\text l}(0,x)-\ds\frac{i}{2}\,\dot \rho(0)\,
g_{\text l}(0,x)\\
\stretch
1&g_{\text l}(0,0)&1\\
\stretch
-\cot\alpha&g'_{\text l}(0,0)&0\endmatrix\right|,\tag 5.6$$
where the overdot denotes the $k$-derivative.

In the exceptional case, we can choose
$$y_1(x)=g_{\text l}(0,x),\quad
y_2(x)=g_{\text l}(0,x)\,\ds\int_0^x \ds\frac{dz}{g_{\text l}(0,z)^2}.$$
In this case, we have
$$y_1(0)=\ds\frac{1+\ell(0)}{\tau(0)},\quad
y'_1(0)=0,\quad y_2(0)=0,\quad y'_2(0)=\ds\frac{1}
{y_1(0)}=\ds\frac{\tau(0)}{1+\ell(0)},$$
with $y_1(0)\ne 0$ because [19,20,22,24] we have $-1<\ell(0)<1.$
Hence, from (5.2) we get
$$\eta(x)=g_{\text l}(0,x)
\left|
\matrix 0&1&\ds\int_0^x \ds\frac{dz}{g_{\text l}(0,z)^2}\\
\stretch
1&g_{\text l}(0,0)&0\\
\stretch
-\cot\alpha&0&\ds\frac{1}{g_{\text l}(0,0)}\endmatrix\right|.\tag 5.7$$

\noindent{\bf Theorem 5.1} {\it The relative
area $[\eta(x)]^2$ for $x\in[0,l]$ is uniquely determined by the
data $\{|F_\alpha(k)|:\ k\in\bR^+\}.$ Equivalently,
$\eta(x)$ for $x\in[0,l]$ is uniquely determined from
$\{Q(x):\ x\in(0,l),\cot\alpha\},$ where $\cot\alpha$
is the constant in (4.9).}

\noindent PROOF: From Theorem~3.1 we know that
$\{|F_\alpha(k)|:\ k\in\bR^+\}$ uniquely determines
the potential $Q(x)$ for $x\in(0,l)$ and
the constant $\cot\alpha.$  From (4.8), (4.9), and (5.3) we see that
$\eta(0)=1$ and $\eta'(0)=-\cot\alpha.$ Thus,
$\eta(x)$ is uniquely obtained by solving the initial value problem
in (5.1) in the special case $A(0)=1.$ \qed

It is possible to construct $\eta(x)$ from $\{|F_\alpha(k)|:\
k\in\bR^+\}$ as follows.
With the help of
(3.4)-(3.12), we can construct
the corresponding right reflection coefficient
$R(k),$ the transmission coefficient $T(k),$ and the Jost solution
$f(k,x).$
Hence, in the generic case
every term appearing on the right
hand side of (5.6) can be constructed from $\{|F_\alpha(k)|:\
k\in\bR^+\},$ and we get
$$\eta(x)=
\left|
\matrix 0&-\ds\frac{i}{2}\,\dot T(0)\,
f(0,x)&i\,\dot f(0,x)-\ds\frac{i}{2}\,\dot R(0)\,
f(0,x)\\
\stretch
1&f(0,0)&1\\
\stretch
-\cot\alpha&f'(0,0)&0\endmatrix\right|.\tag 5.8$$
In the exceptional case, from (5.7) we get
$$\eta(x)=f(0,x)
\left|
\matrix 0&1&\ds\int_0^x \ds\frac{dz}{f(0,z)^2}\\
\stretch
1&f(0,0)&0\\
\stretch
-\cot\alpha&0&\ds\frac{1}{f(0,0)}\endmatrix\right|.$$

Note that we can write the Jost solution $f(k,x)$ as a linear
combination of $g_{\text l}(k,x)$ and $g_{\text l}(-k,x),$ where
$g_{\text l}(k,x)$ is the quantity appearing in (5.5). With the
help of (2.2) we obtain
$$f(k,x)=\ds\frac{e^{ikl}}{2ik}\left|
\matrix 0&g_{\text l}(k,x)&g_{\text l}(-k,x)\\
\stretch
1&g_{\text l}(k,l)&g_{\text l}(-k,l)\\
\stretch
ik&g'_{\text l}(k,l)&g'_{\text l}(-k,l)\endmatrix\right|.\tag 5.9$$
We can also express $F_\alpha(k)$ with the help
of $g_{\text l}(k,x)$. To do so, we can obtain
$f(k,0)$ and $f'(k,0)$ from (5.9) and use
(2.3) to get $F_\alpha(k).$
Alternatively, by using
[cf. (2.5)]
$$g_{\text l}(k,0)=\displaystyle\frac{1+\ell(k)}{\tau(k)},\quad
g'_{\text l}(k,0)=ik\,\displaystyle\frac{1-\ell(k)}{\tau(k)},$$
we can write $F_\alpha(k)$ with
the help of the transmission and
left reflection coefficients associated with $g_{\text l}(k,x).$

\vskip 10 pt
\noindent {\bf 6. RECOVERY FROM THE IMPEDANCE AT THE LIPS}
\vskip 3 pt

The impedance at the lips is defined as
$$Z(k,l):=\ds\frac{p(l,t)}{v(l,t)}.\tag 6.1$$
Using (4.2), (4.11), and (4.18) in (6.1), we get
$$Z(k,l)=\ds\frac{2ick\mu}{2ik\,A(l)+A'(l)}.\tag 6.2$$
Thus, we can only hope to get $A(l)$ and $A'(l)$ from
$Z(k,l).$ We can refer to $Z(k,l)$ as the output impedance
because the volume velocity is from the glottis
to the lips. In Section~11 we will analyze
the impedance at the glottis, which we can
identify as the input impedance.

Note that from (6.2) we get
$$|Z(k,l)|^2=\ds\frac{4c^2k^2\mu^2}{4k^2A(l)^2+A'(l)^2},
\qquad k\in\bR.\tag 6.3$$
By using (6.2) at two distinct real
$k$ values, say $k_1$ and $k_2,$
we can recover $A(l)$ and $A'(l)$ by solving a linear
algebraic system and get
$$A(l)=\ds\frac{c\mu}{k_1-k_2}\left[\ds\frac{k_1}{Z(k_1,l)}-
\ds\frac{k_2}{Z(k_2,l)}\right],
\quad
A'(l)=\ds\frac{2ic\mu k_1k_2}{k_1-k_2}\left[\ds\frac{1}{Z(k_2,l)}-
\ds\frac{1}{Z(k_1,l)}\right].
\tag 6.4$$
On the other hand, if we only know $|Z(k,l)|$ without knowing its
phase, then from (6.3) we get
$A(l)$ and $|A'(l)|$ as
$$A(l)=\sqrt{\ds\frac{c^2\mu^2}{k_1^2-k_2^2}
\left[\ds\frac{k_1^2}{|Z(k_1,l)|^2}-
\ds\frac{k_2^2}{|Z(k_2,l)|^2}\right]},\tag 6.5$$
$$A'(l)^2=\ds\frac{4c^2\mu^2 k_1^2k_2^2}
{k_1^2-k_2^2}\left[\ds\frac{1}{|Z(k_2,l)|^2}-
\ds\frac{1}{|Z(k_1,l)|^2}\right].
\tag 6.6$$
As seen from (6.2), $Z(k,l)$ by itself contains no other information
related to $Q,$ $\eta,$ or $A.$

\vskip 10 pt
\noindent {\bf 7. RECOVERY FROM PRESSURE AT A MICROPHONE}
\vskip 3 pt

Let us place a microphone at a radial distance $r$ from the lips
and measure at that microphone the absolute value of the pressure resulting from
the glottal volume velocity given in (4.1). With the help of (4.2)
it follows that this is equivalent to
having $|P(k,l+r)|$ at hand for $k\in\bR^+.$ From (4.17) we see
that $|P(k,l+r)|$ is an even function of $k\in\bR,$ and from (4.16) we get
$$|P(k,l+r)|^2=
\left(\ds\frac{ck\mu}{4\pi r}\right)^2
\ds\frac{A(l)}{A(0)\,|F_\alpha(k)|^2}
\left[k^2+\,\ds\frac{A'(l)^2}{4\,A(l)^2}\right]
,\qquad k\in\bR.\tag 7.1$$
In this section we show that the data set
$\{|P(k,l+r)|:\ k\in\bR^+\}$ by itself does not uniquely determine
any of $Q(x),$ $\eta(x),$ or $A(x),$ and how additional data
may be used for the unique determination.

\noindent{\bf Theorem 7.1} {\it The data set
$\{|P(k,l+r)|: \ k\in\bR^+,A(l),|A'(l)|\}$
uniquely determines each of $Q(x),$ $\eta(x),$
and $A(x)$ for $x\in(0,l).$}

\noindent PROOF:
It is known (cf. (3.9) of [21]) that for any fixed $\alpha\in(0,\pi)$ we have
$$|F_\alpha(k)|=k+O(1),\qquad k\to+\infty.\tag 7.2$$
Thus, from (7.1) we obtain
$$|P(k,l+r)|=\ds\frac{ck\mu\sqrt{A(l)}}{4\pi r\,\sqrt{A(0)}}\left[1+O(1/k)\right],
\qquad k\to+\infty,$$
or equivalently
$$\sqrt{A(0)}=
\ds\frac{c\mu\sqrt{A(l)}}{4\pi r}\,
\ds\lim_{k\to+\infty}\left[\ds\frac{k}{|P(k,l+r)|}
\right].\tag 7.3$$
Using (7.3) in (7.1) we get
$$\ds\frac{k^2}{|F_\alpha(k)|^2}=
\ds\frac{4\,|P(k,l+r)|^2}{
4k^2+A'(l)^2/A(l)^2}\left[\ds\lim_{k\to+\infty}\ds\frac{k^2}{
|P(k,l+r)|^2}\right],\qquad k\in\bR.\tag 7.4$$
 From (4.17) and (7.4) we see that our data set uniquely determines
$|F_\alpha(k)|$ for $k\in\bR,$ and hence,
as indicated in Theorem~3.1,
$Q(x)$ is uniquely determined for $x\in(0,l).$ Next,
 from Theorem~5.1 it follows that
$\eta(x)$ is also
uniquely determined for $x\in[0,l].$
Finally, from (7.3) we see that
$A(0)$ is also determined by our data set, and thus we recover
$A(x)$ for $x\in[0,l]$ uniquely by using (5.4). \qed

Note that we assume that $A(l)$ and
$|A'(l)|$
do not change with $k,$ and hence they
are constants. As indicated in Section~6
they can be obtained via (6.5) and (6.6) by measuring
the absolute value of the impedance at the lips
at two different frequencies.

\noindent{\bf Theorem 7.2} {\it The data set
$\{|P(k,l+r)|: \ k\in\bR^+,|A'(l)|/A(l)\}$
uniquely determines each of $Q(x)$ and $\eta(x)$
for $x\in(0,l),$ and it determines
$A(x)$ for $x\in[0,l]$ up to a multiplicative constant.}

\noindent PROOF: From (2.4), (4.17), and (7.4) we see
that $|F_\alpha(k)|$ for $k\in\bR$
is uniquely determined by our data set, and hence $Q(x)$ via
Theorem~3.1 and $\eta(x)$
via Theorem~5.1 are uniquely determined for
$x\in(0,l).$ Furthermore, from (3.5) and (4.15) we see
that if we multiply each of
$A(0),$ $A(l),$ and $|A'(l)|$ by the same constant,
we do not change $|p(l+r,t)|$
or equivalently we do not change $|P(k,l+r)|$ for
$k\in\bR.$ Thus, our data set
corresponds to the one-parameter family for $A(x),$ where the parameter
$A(0)$ appears as a multiplicative parameter in (5.4). \qed

With the help of (7.4) and Theorem~7.2 we have the following conclusions.

\noindent{\bf Corollary 7.3} {\it Corresponding to
the data set $\{|P(k,l+r)|: \ k\in\bR^+,A(l)\},$
in general there exists a one-parameter family for each of
$Q(x),$ $\eta(x),$ and $A(x),$ where $|A'(l)|/A(l)$
can be chosen as the parameter.}

\noindent{\bf Corollary 7.4} {\it Corresponding to the
data set $\{|P(k,l+r)|: \ k\in\bR^+\},$
in general there exists a two-parameter family for each of
$Q(x),$ $\eta(x),$ and $A(x),$ where $A(l)$ and
$|A'(l)|$ can be chosen as the parameters.}

\vskip 10 pt
\noindent {\bf 8. RECOVERY FROM THE PRESSURE AT THE LIPS}
\vskip 3 pt

Let us consider the recovery of $A(x)$ for $x\in[0,l]$ from
the absolute value of the pressure at the lips resulting from
the glottal volume velocity in (4.1). From (4.2) we see that our
data set is equivalent to $\{|P(k,l)|:\ k\in\bR^+\}.$ With the help of
(2.4), (2.5), (2.7), and (4.11), we notice that
$|P(k,l)|$ is an even function of $k\in\bR,$ and hence
we have our data actually
available for $k\in\bR.$ In this section we show that this data set uniquely
recovers each of $Q(x),$ $\eta(x),$ and $A(x)$ for $x\in(0,l),$ and
we outline an explicit procedure to determine these quantities.

\noindent{\bf Theorem 8.1} {\it The data set $\{|P(k,l)|: \ k\in\bR^+\}$
uniquely determines each of $Q(x),$ $\eta(x),$ and $A(x)$ for $x\in(0,l).$}

\noindent PROOF: From (2.2), (2.4), and (4.11) we get
$$|P(k,l)|=\ds\frac{c\mu\,|k|}
{\sqrt{A(l)}\,\sqrt{A(0)}\,|F_\alpha(k)|},
\qquad k\in\bR.\tag 8.1$$
Using (7.2) in (8.1), we obtain
$$\sqrt{A(0)\,A(l)}=c\mu
\left(\ds\lim_{k\to+\infty}\left|
\ds\frac{1}{P(k,l)}\right|
\right),\tag 8.2$$
and hence
$$|F_\alpha(k)|=\ds\frac{|k|}
{|P(k,l)|}
\left(\ds\lim_{k\to+\infty}\left|
P(k,l)\right|
\right),\qquad k\in\bR.\tag 8.3$$
As seen from (2.4) and (8.3), by measuring the
absolute value of the pressure at the lips
for $k\in\bR^+,$ we get
$|F_\alpha(k)|$ for $k\in\bR.$
Then, by proceeding as in Section~3,
we can recover $Q(x)$ for $x\in(0,l)$ and the constant $\cot\alpha$
appearing in
(4.9). Next, by proceeding as in Section~5, we determine $\eta(x)$
for $x\in(0,l).$
Note that $\sqrt{A(0)\,A(l)}$ is uniquely determined from our data via (8.2).
Furthermore, as seen from (5.3), we have $\eta(l)=\sqrt{A(l)/A(0)}.$
Thus, we obtain
$$A(0)=\ds\frac{c\mu}{\eta(l)}\left(\ds\lim_{k\to+\infty}\left|
\ds\frac{1}{P(k,l)}\right|
\right),$$
and hence
we get the area function uniquely via (5.4).
\qed

\vskip 10 pt
\noindent {\bf 9. RECOVERY FROM THE TRANSFER FUNCTION}
\vskip 3 pt

The transfer function $\bold T(k,l)$ from the glottis to the lips
is defined as
$$\bold T(k,l):=\ds\frac{v(l,t)}{v(0,t)},\tag 9.1$$
and as we see from (4.1), (4.18), and (9.1), we have
$$\bold T(k,l)=\ds\frac{\sqrt{A(l)}\,e^{-ikl}}{\sqrt{A(0)}\,F_\alpha(-k)}
\left[-k+\ds\frac i2\ds\frac
{A'(l)}{A(l)}\right],\qquad k\in\bCpb.\tag 9.2$$
Hence, with the help of (2.4), we get
$$|\bold T(k,l)|^2=\ds\frac{A(l)}{A(0)\,|F_\alpha(k)|^2}
\left[k^2+\ds\frac
{A'(l)^2}{4A(l)^2}\right],
\qquad k\in\bR.\tag 9.3$$
Using (7.2) in (9.2) we obtain
$$|\bold T(k,l)|=\ds\frac{\sqrt{A(l)}}{\sqrt{A(0)}}\left[1+O(1/k)\right],
\qquad k\to+\infty,$$
and as a result we can recover $A(0)$ as
$$A(0)=\ds\frac{A(l)}{\ds\lim_{k\to+\infty}|\bold T(k,l)|^2}.
\tag 9.4$$
Thus, from (9.2) and (9.4), with the help of (2.4), we have
$$|F_\alpha(k)|^2=\ds\frac{\ds\lim_{k\to+\infty}|\bold T(k,l)|^2}
{|\bold T(k,l)|^2}\left[k^2+\ds\frac14\ds\frac{A'(l)^2}{A(l)^2}
\right],\qquad k\in\bR.$$

Comparing (9.3) with (7.1) we see that
$$|\bold T(k,l)|^2=
\left(\ds\frac{4\pi r}{ck\mu}\right)^2|P(k,l+r)|^2
,\qquad k\in\bR,$$
and hence we have the following conclusion.

\noindent{\bf Corollary 9.1} {\it For each fixed $r>0,$
the information contained in the data set
$\{|\bold T(k,l)|:\ k\in\bR^+\}$ is equivalent to that in
$\{|P(k,l+r)|:\ k\in\bR^+\}.$}

In other words, measuring the absolute value of the pressure
at a microphone placed at some distance
 from the lips is equivalent to measuring the absolute value of
the transfer function from the glottis to the lips.
Consequently, we have the following analogs of the results of Section~7.

\noindent{\bf Corollary 9.2} {\it The data set
$\{|\bold T(k,l)|: \ k\in\bR^+,A(l),|A'(l)|\}$
uniquely determines each of $Q(x),$ $\eta(x),$
and $A(x)$ for $x\in(0,l).$}

\noindent{\bf Corollary 9.3} {\it The data set
$\{|\bold T(k,l)|: \ k\in\bR^+,|A'(l)|/A(l)\}$
uniquely determines each of $Q(x)$ and $\eta(x)$
for $x\in(0,l)$ and it determines
$A(x)$ for $x\in[0,l]$ up to a multiplicative constant.}

\noindent{\bf Corollary 9.4} {\it Corresponding to
the data set $\{|\bold T(k,l)|: \ k\in\bR^+,A(l)\},$
in general there exists a one-parameter family for each of
$Q(x),$ $\eta(x),$ and $A(x),$ where $|A'(l)|/A(l)$
can be chosen as the parameter.}

\noindent{\bf Corollary 9.5} {\it Corresponding to the data set
$\{|\bold T(k,l)|: \ k\in\bR^+\},$
in general there exists a two-parameter family for each of
$Q(x),$ $\eta(x),$ and $A(x),$ where $A(l)$ and $|A'(l)|$
can be chosen as the parameters.}

\vskip 10 pt
\noindent {\bf 10. RECOVERY FROM A GREEN'S FUNCTION AT THE LIPS}
\vskip 3 pt

In this section we show that the absolute
value of a Green's function for (1.3) at the lips measured for
$k\in\bR^+$
enables us to uniquely construct each of
$Q(x),$ $\eta(x),$ and $A(x)$ for $x\in(0,l).$

The Green's function at the lips can be defined [17] as the solution
$\Phi(l,t)$ given in (1.2) when
the glottal volume velocity is as in (4.1).
Thus, from (1.2), (4.2), and (4.11), we get the Green's function at the lips
as
$$\bold G(k,l;t)= \ds\frac{-ck\mu\,e^{ik(ct-l)}}
{\sqrt{A(0)}\,F_\alpha(-k)}.$$
Hence, with the help of (2.4) we obtain
$$|\bold G(k,l;t)|=\ds\frac{c\,|k|\,\mu}
{\sqrt{A(0)}\,|F_\alpha(k)|},\qquad k\in\bR.\tag 10.1$$

\noindent {\bf Theorem 10.1} {\it The data set $\{|\bold G(k,l;t)|:\ k\in\bR^+\}$
uniquely determines each of $Q(x),$ $\eta(x),$ and $A(x)$ for $x\in(0,l).$}

\noindent PROOF:
 From (2.4) and (10.1) it follows that
$|\bold G(k,l;t)|$ is independent of $t$ and
is an even function of $k$ on $\bR,$
and hence our data can be extended from $k\in\bR^+$ to $k\in\bR.$
Using (7.2) in (10.1) we get
$$\sqrt{A(0)}=\ds\frac{c\mu}
{\ds\lim_{k\to+\infty}\left|\bold G(k,l;t)\right|},\tag 10.2$$
and hence
$$\frac{|k|}{|F_\alpha(k)|}=
\ds\frac{|\bold G(k,l;t)|}{\ds\lim_{k\to+\infty}|\bold G(k,l;t)|}
,\qquad k\in\bR.$$
Thus, we get $|F_\alpha(k)|$ for $k\in\bR$ whenever
we have $|\bold G(k,l;t)|$ for $k\in\bR^+.$ Then, as in Section~3
we construct $Q(x)$ for
$x\in(0,l)$ and $\cot\alpha.$ Next, as in Section~5, we construct
$\eta(x)$ for $x\in[0,l].$ Finally, with the help of
(5.4) and (10.2) we obtain
$$A(x)=\ds\frac{c^2\mu^2\,[\eta(x)]^2}
{\left(\ds\lim_{k\to+\infty}\left|\bold G(k,l;t)\right|\right)^2}.$$
Thus, the proof is complete. \qed

\vskip 10 pt
\noindent {\bf 11. RECOVERY FROM THE IMPEDANCE AT THE GLOTTIS}
\vskip 3 pt

The impedance at the glottis is defined as
$$Z(k,0):=\ds\frac{p(0,t)}{v(0,t)}.\tag 11.1$$
Using (4.1), (4.2), and (4.11) in (11.1) we get
$$Z(k,0)=-\ds\frac{ck\mu\,f(-k,0)}{A(0)\,F_\alpha(-k)}.\tag 11.2$$
 From (2.4), (2.5), (2.7), and (11.2) we see that $|Z(k,0)|$ is an even
function of $k\in\bR,$ and hence $|Z(k,0)|$
is known for $k\in\bR$ if it is known for
$k\in\bR^+.$
In this section we show that the information contained in
$\{|Z(k,0)|:\ k\in\bR^+\}$ enables us to
uniquely construct $Q,$ $\eta,$ and $A.$

\noindent {\bf Theorem 11.1} {\it The data set $\{|Z(k,0)|:\ k\in\bR^+\}$
uniquely determines each of $Q(x),$ $\eta(x),$ and $A(x)$ for $x\in(0,l).$}

\noindent PROOF: With the help of (2.4) and (11.2) we get
$$|Z(k,0)|=\ds\frac{c\,|k|\,\mu\,|f(k,0)|}{A(0)\,|F_\alpha(k)|},
\qquad k\in\bR,\tag 11.3$$
and hence, using (7.2) and the fact [18-20,22] that
$f(k,0)=1+O(1/k)$ as $k\to+\infty,$ we obtain
$$|Z(k,0)|=\ds\frac{c\mu}{A(0)}\left[1+O(1/k)\right],\qquad
k\to+\infty,$$
which leads to
$$A(0)=\ds\frac{c\mu}{\ds\lim_{k\to+\infty}|Z(k,0)|}.\tag 11.4$$
Thus, we can write (11.3) as
$$\left|\ds\frac{k\,f(k,0)}{F_\alpha(k)}\right|=\ds\frac{|Z(k,0)|}
{\ds\lim_{k\to +\infty}|Z(k,0)|}
.\tag 11.5$$
As seen from (11.5), the recovery from
$|Z(k,0)|$ for $k\in\bR^+$ is equivalent
to the recovery from $|f(k,0)/F_\alpha(k)|$ for
$k\in\bR.$

Recall that we assume that the half-line
Schr\"odinger equation with the boundary condition (2.1)
does not have any bound states. In that case,
it is known [21] that $kf(k,0)/F_\alpha(k)$ is
analytic in $\bCp,$ continuous in $\bCpb,$ nonzero
in $\bCpb\setminus\{0\},$ and it is either
nonzero at $k=0$ or has a simple zero there, and
$$\ds\frac{k\,f(k,0)}{F_\alpha(k)}=1+O(1/k),\qquad
k\to\infty\text{ in } \bCpb.$$
As a result, we can recover $kf(k,0)/F_\alpha(k)$ for
$k\in\bCpb$ from its amplitude
known for $k\in\bR$ via
$$\ds\frac{k\,f(k,0)}{F_\alpha(k)}=\exp\left(
\ds\frac{1}{2\pi}\int_{-\infty}^\infty
dt\,\ds\frac{\log|t\,f(t,0)/F_\alpha(t)|}{
t-k-i0^+}\right),\qquad
k\in\bCpb.$$
Having constructed $kf(k,0)/F_\alpha(k)$ for
$k\in\bR,$ we can use (3.2) and obtain
$\Lambda_\alpha(k)$ for $k\in\bR.$ Next, using (3.4)
we construct $|F_\alpha(k)|$ for $k\in\bR.$ Then, as in Section~3
we construct $Q(x)$ for $x\in(0,l),$ and as in Section~5
we construct $\eta(x)$ for $x\in[0,l].$ Finally, $A(x)$ is constructed
via (5.4) after obtaining $A(0)$ from (11.4). \qed

\vskip 10 pt
\noindent {\bf 12. RECOVERY FROM THE REFLECTANCE AT THE GLOTTIS}
\vskip 3 pt

The reflectance at the glottis is defined [3,4] as the ratio of
the right-moving (from the glottis towards the lips) pressure wave to the left-moving
pressure wave. From (4.11) that reflectance
is seen to be equal to $L(-k)$ because the Jost solution to the
full-line Schr\"odinger equation has the extension
$$f(k,x)=\displaystyle\frac{1}{T(k)}\,e^{ikx}+
\displaystyle\frac{L(k)}{T(k)}\,e^{-ikx}
,\qquad x\le 0,$$
where we recall that
$L(k)$ is the left
reflection coefficient appearing in (2.5).
In this section we show that either of the real and
imaginary parts of the reflectance at the glottis
known for $k\in\bR^+$ enables us
to construct $Q(x)$ uniquely for $x\in(0,l).$ On the other hand,
it cannot uniquely determine either $\eta(x)$ or $A(x).$

\noindent {\bf Theorem 12.1} {\it Either of the data sets
$\{\text{\rm Re}[L(k)]:\ k\in\bR^+\}$ and $\{\text{\rm Im}[L(k)]:\ k\in\bR^+\}$
uniquely determines $Q(x)$ for
$x\in(0,l).$ On the other hand, to each of these data sets, there
correspond a one-parameter family for $\eta(x)$ and
a two-parameter family for $A(x).$}

\noindent PROOF:
 From (2.7) we see that $\text{Re}[L(k)]$ is an even function of
$k$ on $\bR$ and $\text{Im}[L(k)]$ is an odd function. Hence, both of
our data sets can be extended from $k\in\bR^+$ to $k\in\bR.$
In the absence of bound states,
because $Q(x)\equiv 0$ for $x<0,$ it is known [22,24] that $L(k)$ is analytic
in $\bCp,$ continuous in $\bCpb,$ and $o(1/k)$ as $k\to\infty$
in $\bCpb.$ Thus, we can construct $L(k)$ from either data set
via the Schwarz integral formula [cf. (3.4)] as
$$L(k)=\ds\frac{1}{\pi i}
\int_{-\infty}^\infty
\ds\frac{dt\, \text{Re}[L(t)]}{t-k-i0^+}=
\ds\frac{1}{\pi}
\int_{-\infty}^\infty
\ds\frac{dt\, \text{Im}[L(t)]}{t-k-i0^+}
,\qquad k\in\bCpb.$$
We can then obtain $Q$ uniquely, for example, via [22,24]
the Faddeev-Marchenko method, as
$$Q(x)=2\,\ds\frac{d}{dx}B_{\text r}(x,0^+),\qquad x\in\bR,$$
where
$B_{\text r}(x,y)$ is obtained by solving the (right)
Faddeev-Marchenko integral equation
$$B_{\text r}(x,y)+\hat L(-2x+y)+\int_0^\infty dz\,
\hat L(-2x+y+z)\,B_{\text r}(x,z)=0,\qquad
x\in\bR, \quad y>0.$$
It is already known [19,20,22,24] that $T(k)$ and $R(k)$
can be constructed from $L(k)$ with the help of
(2.6) and
$$T(k)=\exp\left(\displaystyle\frac{1}{2\pi i}\int_{-\infty}^\infty
ds\,\displaystyle\frac{\log(1-|L(s)|^2)}
{s-k-i0^+}\right), \qquad
k\in\overline{\bold C^+},$$
and hence, as we see from (5.8), $\eta(x)$ is not
uniquely determined and we have the corresponding one-parameter
family for $\eta(x)$ with $\cot\alpha$ being the parameter. From
(5.1) we see that we have the corresponding two-parameter family for
$A(x),$ where the parameters can be chosen, for example, as
$A(0)$ and $A'(0),$ or as $A(l)$ and $A'(l).$ \qed

Finally, let us remark that, as indicated in (6.4), we can obtain
$A(l)$ and $A'(l)$ if we know the impedance at the lips at
two distinct real $k$ values.

\vskip 10 pt
\noindent {\bf 13. EXAMPLES}
\vskip 3 pt

In this section we illustrate the theoretical results presented in the
previous sections with some examples.

Let us use $l=17.5$ cm,
$c=3.43\times 10^4$ cm/sec,
$\mu=1.2\times 10^{-3}$ gm/cm$^3,$ $A(0)=5$ cm$^2$, and
$A'(0)=-0.52$ cm, and
$$Q(x)=\ds\frac{80(7 + 3\sqrt{5})\,e^{2\sqrt{5}x}}
{\left[(7 + 3\sqrt{5})\,e^{2\sqrt{5}x}-2\right]^2}.$$
When $Q$ is viewed as a potential of the full-line
Schr\"odinger equation
with support on $\bR^+,$ the
corresponding scattering coefficients
$\tau(k),$ $\rho(k),$ $\ell(k)$
and the left Jost solution $g_{\text l}(k,x)$
introduced in Section~5
are rational functions of $k$, and it can be verified that
$$g_{\text l}(k,x)=
e^{ikx}\left[1+\ds\frac{i}{k+i\sqrt{5}}
\,\ds\frac{4i\sqrt{5}}{(7 + 3\sqrt{5})\,e^{2\sqrt{5}x}-2}\right],
\qquad x\ge 0,$$
$$\tau(k)=\ds\frac{k(k+i\sqrt{5})}{(k+i)(k+2i)},
\quad \ell(k)=\ds\frac{2}{(k+i)(k+2i)}
,$$
$$
\rho(k)=\ds\frac{-2(k+i\sqrt{5})}{(k+i)(k+2i)(k-i\sqrt{5})}.$$
All the quantities related to (1.1), (1.3), and (1.6)
can now be explicitly evaluated. For example,
the left Jost solution $g_{\text l}(k,x)$ for $x\le 0$ for the full-line
Schr\"odinger equation can be obtained as
$$g_{\text l}(k,x)=\ds\frac{e^{ikx}}{\tau(k)}+\ds\frac{\ell(k)\,e^{-ikx}}
{\tau(k)},\qquad x\le 0.$$
Via (4.9) we get $\cot\alpha=-0.052,$ $\eta(x)$ can be obtained via (5.6),
$A(x)$ via (5.4), $f(k,x)$ via (5.9), $F_\alpha(k)$ via (2.3),
the scattering coefficients $T(k),$ $R(k),$ and $L(k)$ via (3.9) and
(3.10), $P(k,x)$ via (4.11), $v(x,t)$ via (4.18),
$|P(k,l+r)|$ via (7.1), $|P(k,l)|$ via (8.1),
$|Z(k,l)|$ via (6.3), $|\bold T(k,l)|$ via (9.3),
$|\bold G(k,l;t)|$ via (10.1),
$|Z(k,0)|$ via (11.3),
$\Lambda_\alpha(k)$ via (3.2). We also compute
$A(l)=11.596$ cm$^2$ and $A'(l)=0.681$ cm. Even though all these
quantities can be explicitly written in terms of elementary functions
in closed forms,
the corresponding expressions are too long to display here, and instead
we only show some of their graphs.

\vskip 15 pt

\centerline{\hbox{\psfig{figure=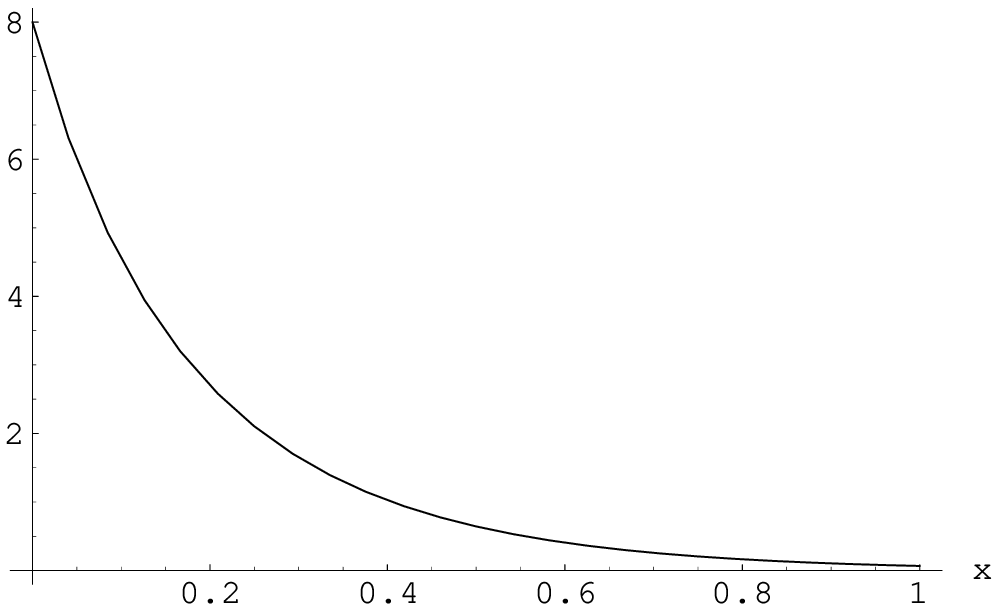,height=1.3 truein,width=1.9 truein}}
\ {\psfig{figure=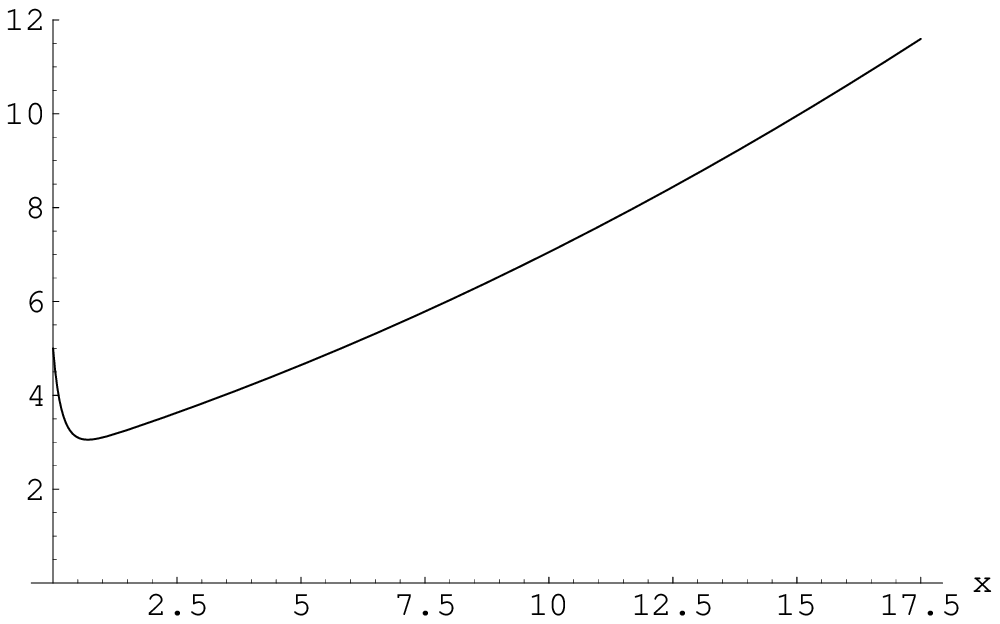,height=1.3 truein,width=1.9 truein}}
\ {\psfig{figure=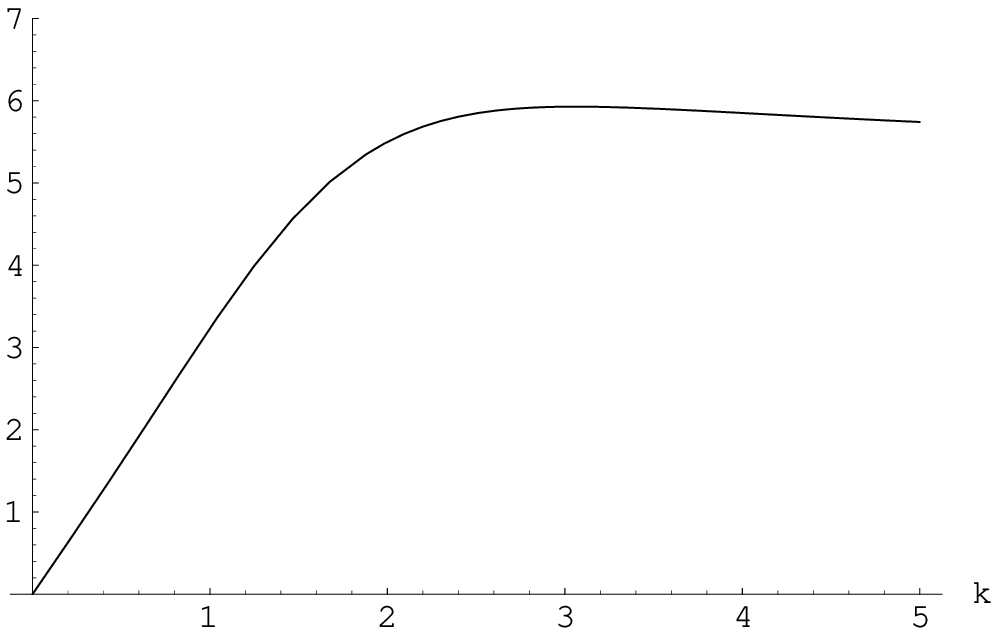,height=1.3 truein,width=1.9 truein}}}

\centerline{\quad {\bf Fig. 13.1} \ $Q(x).$ \qquad \qquad \quad
{\bf Fig. 13.2} $A(x).$ \qquad \qquad{\bf Fig. 13.3} \
$|P(k,l)|.$}

\vskip 15 pt

\centerline{\hbox{\psfig{figure=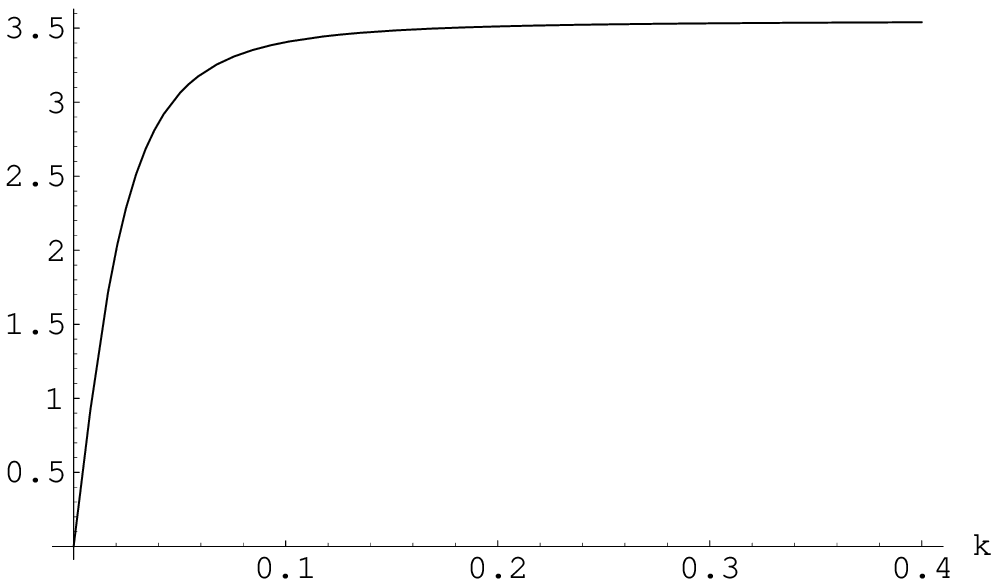,height=1.3 truein,width=1.9 truein}}
\ {\psfig{figure=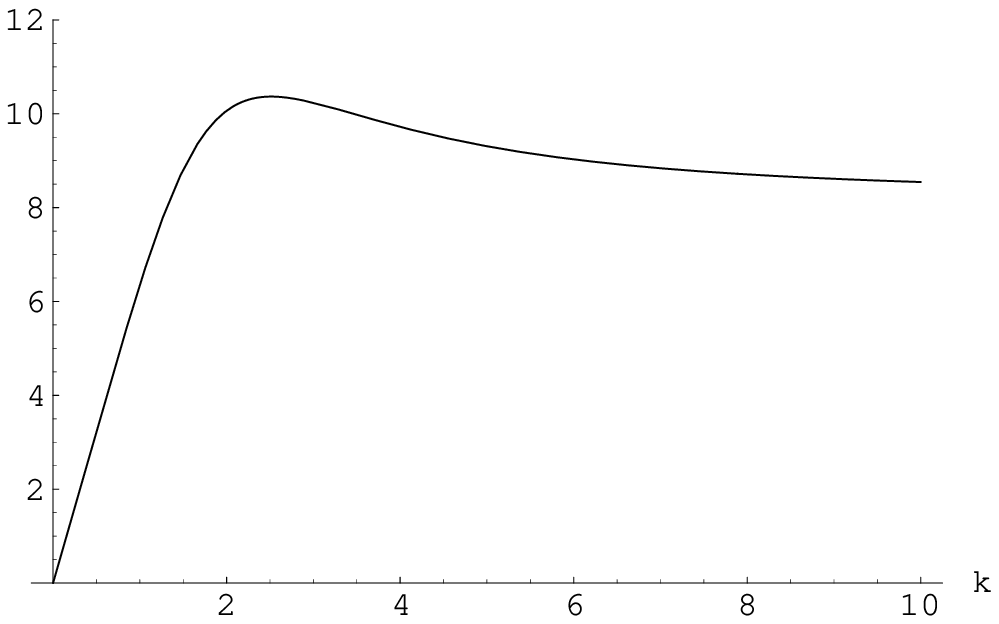,height=1.3 truein,width=1.9 truein}}
\ {\psfig{figure=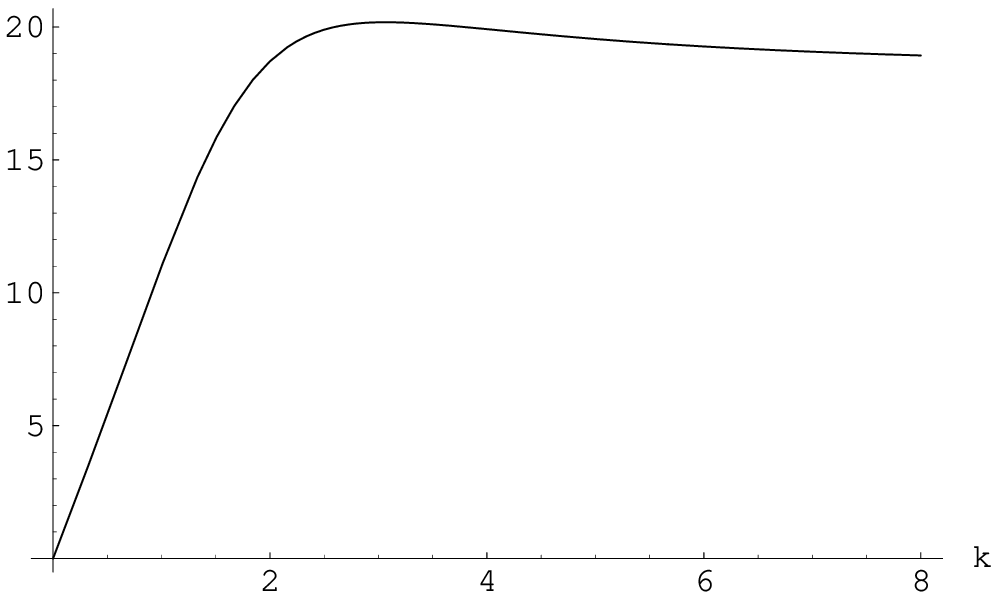,height=1.3 truein,width=1.9 truein}}}

\centerline{\qquad {\bf Fig. 13.4} \  $|Z(k,l)|.$ \qquad \quad
{\bf Fig. 13.5} \ $|Z(k,0)|.$ \qquad
{\bf Fig. 13.6} \ $|\bold G(k,l;t)|.$}

\vskip 15 pt

\centerline{\hbox{\psfig{figure=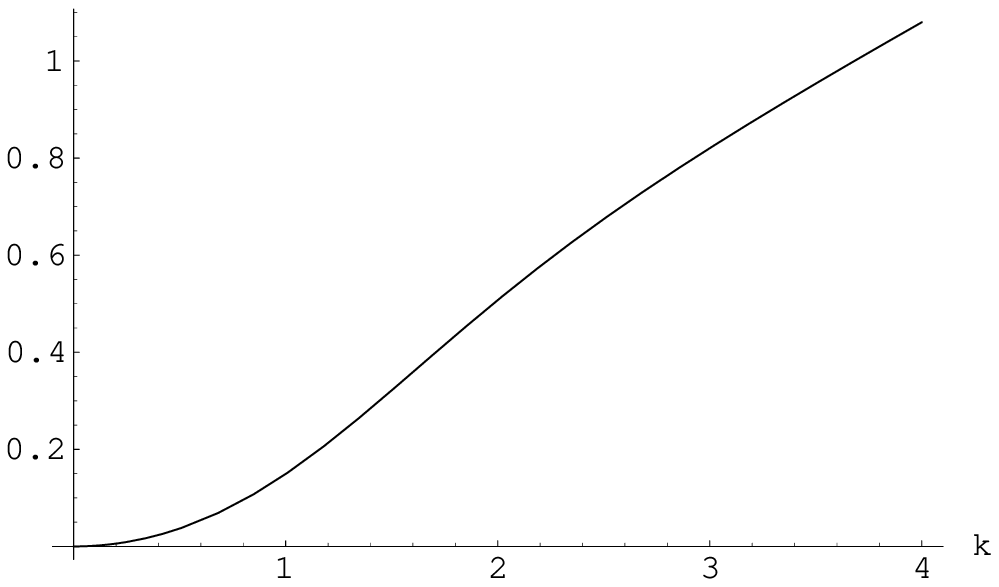,height=1.3 truein,width=1.9 truein}}
\ {\psfig{figure=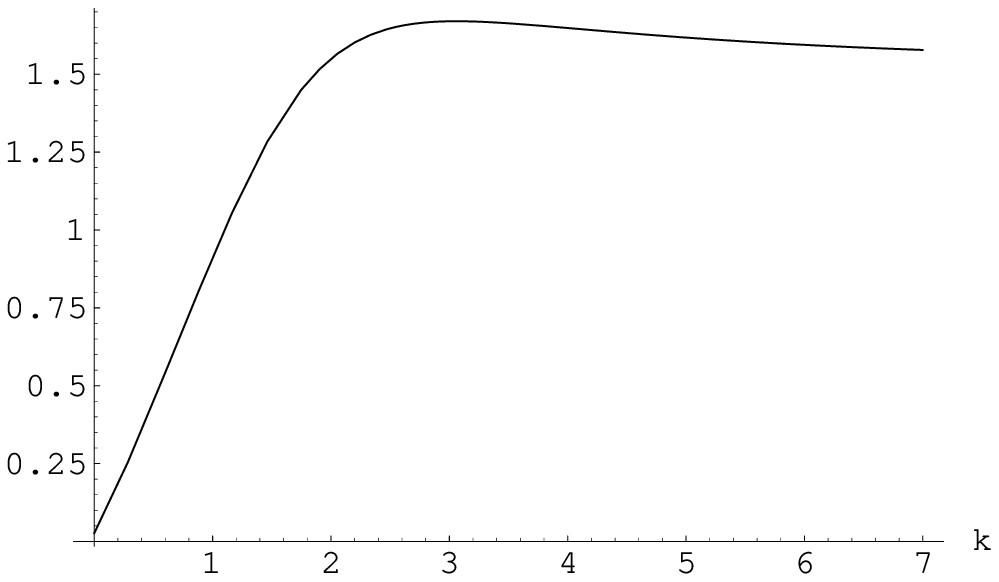,height=1.3 truein,width=1.9 truein}}
\ {\psfig{figure=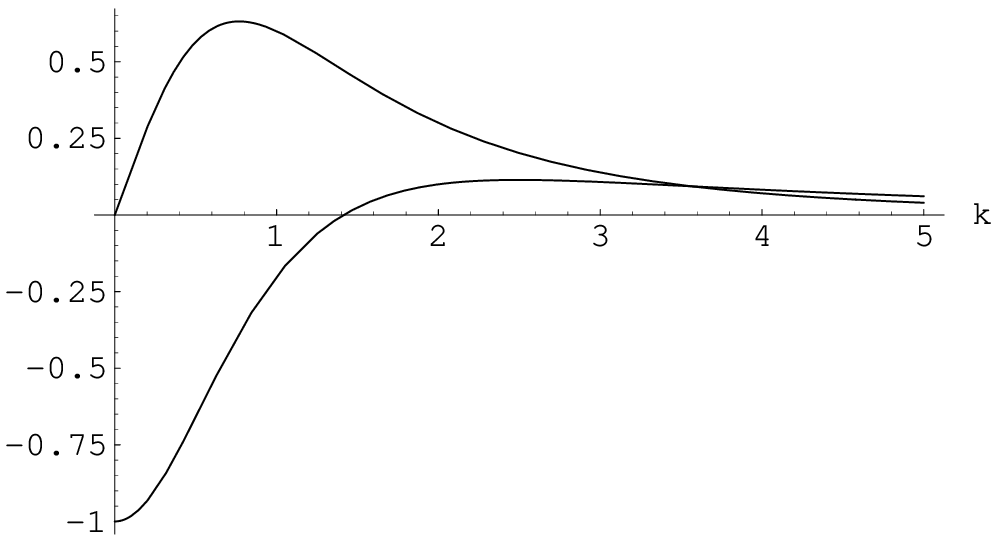,height=1.3 truein,width=1.9 truein}}}

\centerline{{\bf Fig. 13.7} \ $|P(k,l+20)|.$
\qquad
{\bf Fig. 13.8} \ $|\bold T(k,l)|.$
\qquad \quad
{\bf Fig. 13.9} \ $L(-k).$}

Let us remark that in Fig.~13.7 the absolute value
of the pressure is measured at a distance of 20 cm from the
lips. In Fig.~13.9 the real and imaginary parts of the reflectance
are displayed, which can be told apart from
$\text{Re}[L(0)]=-1$ and $\text{Im}[L(0)]=0.$

It is known from Section~6 that the graph in Fig.~13.4 cannot determine
either of Figs.~13.1 and 13.2. We know from Section~8 that the graph
in Fig~13.3 uniquely determines the graphs of Figs.~13.1 and 13.2.
As we know from Section~11, the graph
in Fig~13.5 uniquely determines the graphs of Figs.~13.1 and 13.2.
 From Section~10 we know that the graph in
Fig~13.6 also uniquely determines the graphs of Figs.~13.1 and 13.2.
We know from Section~9 that the graphs of Figs.~13.7 and 13.8 contain
the same information, but neither is sufficient to determine uniquely
either of the graphs in Figs.~13.1 and 13.2. On the other hand, as we
have seen in Sections~7 and 9, the graphs in Figs.~13.4 and 13.7
together, or the graphs in Figs.~13.4 and 13.8 together,
uniquely determine the graphs of
Figs.~13.1 and 13.2. We know from Section~12 that either graph in
Fig.~13.9 uniquely constructs the graph in Fig.~13.1, but not that in Fig.~13.2.

\vskip 10 pt

\noindent{\bf Acknowledgments.}
The author has benefited from discussions
with Roy Pike, Barbara Forbes, and Paul Sacks.
The research leading to this article
was supported in part by the National Science Foundation
under grant DMS-0204437 and
the Department of Energy under grant
DE-FG02-01ER45951.

\vskip 5 pt

\noindent {\bf{REFERENCES}}

\item{[1]} G. Fant, {\it Acoustic theory of speech production,} Mouton,
The Hague, 1970.

\item{[2]} J. L. Flanagan, {\it Speech analysis synthesis and perception,}
2nd ed., New York, Springer, 1972.

\item{[3]} M. M. Sondhi, {\it A survey of the vocal tract inverse problem:
theory, computations and experiments,} in: F. Santosa, Y. H. Pao,
W. W. Symes, and C. Holland (eds.),
{\it Inverse problems of acoustic and elastic waves,} SIAM,
 Philadelphia, 1984, pp. 1--19.

\item{[4]} J. Schroeter and M. M. Sondhi, {\it Techniques
for estimating vocal-tract shapes from the speech signal,}
IEEE Trans. Speech Audio Process. {\bf 2}, 133--149 (1994).

\item{[5]} K. N. Stevens, {\it Acoustic phonetics,} MIT Press,
Cambridge, MA, 1998.

\item{[6]} G. Borg, {\it Eine Umkehrung der Sturm-Liouvilleschen
Eigenwertaufgabe,} Acta Math. {\bf 78}, 1--96 (1946).

\item{[7]} M. R. Schroeder,
{\it Determination of the geometry of the human vocal tract by
acoustic measurements,} J. Acoust. Soc. Am. {\bf 41}, 1002--1010 (1967).

\item{[8]} P. Mermelstein, {\it
Determination of the vocal-tract shape from measured formant frequencies,}
J. Acoust. Soc. Am. {\bf 41}, 1283--1294 (1967).

\item{[9]}
B. Gopinath and M. M. Sondhi,
{\it Determination of the shape of the human vocal tract
shape from acoustical measurements,}
Bell Sys. Tech. J. {\bf 49}, 1195--1214 (1970).

\item{[10]} L G\aa rding, {\it The inverse of vowel articulation,}
Ark. Mat. {\bf 15}, 63--86 (1977).

\item{[11]} J. R. McLaughlin,
{\it Analytical methods for recovering coefficients in
differential equations from spectral data,}
SIAM Rev. {\bf 28}, 53--72 (1986).

\item{[12]}
M. M. Sondhi and B. Gopinath,
{\it Determination of vocal-tract shape from impulse response
at the lips,} J. Acoust. Soc. Am. {\bf 49}, 1867--1873 (1971).

\item{[13]} R. Burridge,
{\it The Gelfand-Levitan, the Marchenko, and the
Gopinath-Sondhi integral equations of inverse scattering
theory, regarded in the context of inverse impulse-response
problems,} Wave Motion {\bf 2}, 305--323 (1980).

\item{[14]} M. M. Sondhi and
J. R. Resnick,
{\it The inverse problem for the vocal tract:
numerical methods, acoustical experiments, and speech synthesis,}
J. Acoust. Soc. Am. {\bf 73},
985--1002 (1983).

\item{[15]} W. W. Symes, {\it On the relation between coefficient
and boundary values for solutions of Webster's Horn equation,}
SIAM J. Math. Anal. {\bf 17}, 1400--1420 (1986).

\item{[16]} Rakesh,
{\it Characterization of transmission data for Webster's horn equation,}
Inverse Problems {\bf 16}, L9--L24 (2000).

\item{[17]} B. J. Forbes, E. R. Pike, and D. B. Sharp,
{\it The acoustical Klein-Gordon equation: The wave-mechanical
step and barrier potential functions,} J. Acoust. Soc. Am. {\bf
114}, 1291--1302 (2003).

\item{[18]} I. M. Gel'fand and B. M. Levitan,
{\it On the determination of a differential equation from its
spectral function,}
Am. Math. Soc. Transl. (ser. 2) {\bf 1}, 253--304 (1955).

\item{[19]}  V.\; A.\; Marchenko, {\it Sturm-Liouville operators and
applications,} Birk\-h\"au\-ser, Basel, 1986.

\item{[20]} B. M. Levitan, {\it Inverse Sturm-Liouville problems,}
VNU Science Press, Utrecht, 1987.

\item{[21]} T. Aktosun, and R. Weder,
{\it Inverse spectral-scattering problem with two sets of
discrete spectra for the radial Schr\"odinger equation,}
IMA preprint \#1960, (2004).

\item{[22]} K. Chadan and P. C. Sabatier, {\it Inverse problems in quantum
scattering theory,} 2nd ed., Springer, New York, 1989.

\item{[23]} K. Chadan and P. C. Sabatier, {\it Chapter 2.2.1,
Radial inverse scattering problems,} in: E.
R. Pike and P. C. Sabatier (eds.), {\it Scattering,} Academic
Press, London, 2001, pp. 726--741.

\item{[24]} T. Aktosun and M. Klaus, {\it
Chapter 2.2.4, Inverse theory: problem on the line,} in: E. R.
Pike and P. C. Sabatier (eds.), {\it Scattering,} Academic Press,
London, 2001, pp. 770--785.

\item{[25]} T. Aktosun, {\it Construction of the half-line
potential from the Jost function,}
Inverse Problems {\bf 20}, 859--876 (2004).

\end